# Towards Specialized Supercomputers for Climate Sciences: Computational Requirements of the Icosahedral Nonhydrostatic Weather and Climate Model

Torsten Hoefler[1], Alexandru Calotoiu[1], Anurag Dipankar[1], Thomas Schulthess[1], Xavier Lapillonne[2], Oliver Fuhrer[2]

[1] ETH Zurich; [2] MeteoSwiss

While the impact of climate change is clearly visible in our daily experiences, news, and short-term predictions, we are not yet equipped to predict long-term effects at accurate spatial and temporal resolutions months, years, or decades in the future. For example, while all simulations agree that the global-mean temperature is increasing, they are starkly disagreeing on the exact dynamics of this increase. One or two degrees may make the difference, especially near points that may fundamentally change ecosystems (Gruber et al. 2020). An important simulation 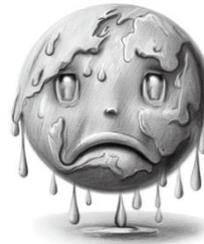 scenario is the Equilibrium Climate Sensitivity (ECS), which estimates the global average temperature increase given that the $CO_2$ concentration doubled. Predictions for ECS range from 2-5°C and the uncertainty is mainly caused by different representations of clouds (Forster, 2021). This global average prediction is likely the simplest important metric of $CO_2$'s impact and acts as a good litmus test to see where the quality of climate models stands. If we cannot predict average global temperature change with high confidence, we will surely not be able to predict the frequency or likelihood of localised events.

Yet, such localised predictions of likelihood of future extreme events are crucial to drive prevention as well as adaptation. The most promising avenue to improve this prediction is to increase the model's resolution to capture the physics of clouds and global circulation using first principles. This also enables **more accurate predictions of future climate scenarios in specific regions, which can guide our decisions where to invest in local infrastructure to mitigate extreme events**. Today's simulations operate at resolutions of tens of kilometres and are mainly limited by our available compute infrastructure. Large thunderstorm clouds are resolved at single-digit km-scale while smaller cloud formations need hundreds of metres resolution. Clouds have a large impact on the Earth's climate, not only through precipitation events but also reflection or absorption of radiation originating from the sun or reflected by the surface. Thus, to enable accurate predictions, we need to push climate simulations to higher resolution. Unfortunately, each doubling in horizontal resolution of a climate simulation increases the computational requirements between 8-16 times. Given where we stand today, **we require 3-4 orders of magnitude higher performance to achieve this goal**.

To achieve these orders of magnitude we need to **push future computer architecture and software design to their limits to achieve higher performance**. In this work, we make a step towards this goal: **we analyse the performance of a leading complex climate simulation code with more than 630k lines of Fortran code in detail**. This model will not only help us to understand the compute and storage requirements of a candidate high-resolution simulation in detail, but it may also uncover scaling issues in the components themselves and guide code optimization efforts to "where the puck is going". Furthermore, our requirements models return the exact number of floating-point operations as well as information about data movement. Thus, **our requirements models can guide the design of next-generation supercomputers to optimally execute km-scale climate simulations**.

The **Icosahedral Nonhydrostatic Weather and Climate Model (ICON)** is a complex modelling framework designed to capture all effects needed to accurately model the Earth. It is a software maintained by the community and led by a collaboration between the German Weather Service (DWD) and the Max Planck Institute for Meteorology and has hundreds of users and contributors aiming to become a unified global numerical weather and climate model. One of the main goals for the collaboration is to develop an accurate model code that scales to massively parallel supercomputers to achieve kilometre-scale resolution for global and regional forecasting (Prill et al, 2023).

As a **climate and weather modelling framework**, ICON supports a wide range of models, such as ocean, atmosphere, land, etc., that can be **composed into configurations**. For example, numerically predicting the regional weather will not need to model oceans like a climate simulation model but may require modelling different physical effects (e.g., pollen). Thus, the large number of components in ICON can be **configured into a runnable model like Lego building blocks** can be combined to model both a plane or a car.

We design a methodology to quickly derive performance models that capture the requirements as well as execution performance properties of specific ICON configurations. While these models differ for each configuration at hand, we follow ICON's philosophy and timer infrastructure to develop an easy-to-use performance modelling framework for all conceivable ICON configurations.

This framework enables us to conduct **back-of-the envelope** calculations for a large-scale **30-year climate simulation run at high (1.23 km) resolution with ICON**. We show that this run (in the current implementation and configuration we study) **requires about 11 Zflop**. If we assume an optimistic hypothetical floating-point efficiency of 1% on an **exaflop/s system, we would require about 13 days of the full system to generate one simulation trajectory** (at about 2.4 Simulated Years Per Day (SYPD) throughput). Running with $2^{14}$ ($\approx 16k$) MPI processes, the total communication volume at each process would be 3.66 PiB, making an average communication bandwidth of 3.25 GiB/s per process. Assuming we output seven 3D variables at 70 layers in 16-bit precision every six hours of simulated time, we **would require 14.4 PiB of storage**! This would equal to an **average I/O bandwidth of 12.8 GiB/s** to a file system. We see a huge potential to compress the output before writing (Huang et al., 2023, Kloewer et al., 2021). Assuming we run the job with 16k MPI processes, our model predicts a **problematic memory requirement per process of 95.7 GiB**. We suspect a memory scalability bug in the MPI communication parts of the code.

This is just one simple example of how the models can be used to derive a system design with specific network and file system bandwidths and floating-point performance numbers. More advanced uses of the model enable programmers to focus on which components become important at scale. One major insight is that **the relative importance of the dynamical core quickly grows relative to the physical parameterization due to the finer time-stepping**. Thus, optimizations for large-scale runs should focus on the dynamical core.

**3.1 Towards Machine-Learning Acceleration**

Artificial intelligence-based techniques will soon play an important role to improve climate simulations in speed and accuracy. Data-driven statistical machine learning predictions can replace pieces of a simulation ("ML inside"), replace full simulations, or post-processing and analyses ("ML on top") (Hoefler et al., 2023, Bauer et al., 2023). While this paper does not directly consider these techniques, it addresses the most important gap towards enabling them: data to learn from. Today, **the lack of high-resolution data to train machine learning models is the biggest impediment to having accurate machine learning models**. Many models train at a coarse resolution of 0.25 degree (27.8 km at the equator), the resolution of the ERA-5 dataset (Hersbach et al, 2020). Unfortunately, 0.25-degree resolution is not sufficient to model many important cloud phenomena (Hoefler et al., 2023). In this work, we outline the computational requirements for producing high-resolution data that is needed to train high-precision climate machine learning models. We also address storage requirements while we note that the data could be streamed through a learning model without being stored or it could be compressed significantly.

Furthermore, our performance models of a realistic production-level weather and climate code will enable researchers and practitioners to focus their attention to the most promising pieces for supplementation by artificial intelligence modules. Furthermore, the ICON-specific model can easily be generalised to other applications using a similar methodology. Thus, **our performance modelling work forms an important first step in the long journey towards high-resolution machine learning predictions in weather and climate sciences**.

### 3.2 ICON Grids

ICON discretizes the simulated domain using an icosahedral grid that can tile the spherical Earth relatively uniformly. Each ICON run requires a grid to model the two surface (horizontal) dimensions of the simulation that can be generated with the tool icongridgen. Icosahedral grids in ICON are created recursively by starting from a

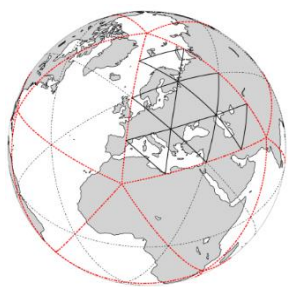

**Source: ICON Tutorial 2023**

convex regular icosahedron with 20 faces and dividing the edges into *n* parts followed by *k* iterative edge bisections --- the resulting grid is called R*n*B*k*.

The figure on the right (Prill et al, 2023) illustrates the construction of a R2B1 grid: the original icosahedron is in red; the single edge division is shown in dotted black lines; and the final single bisection is shown with solid red lines. The number of grid points for an ICON mesh is thus $N = 20n^2\, 4^k$ and the grid resolution is approximately $\Delta x \approx 5050/(n \cdot 2^k) km$. The resulting grid is largely regular: by construction each newly created point has six neighbours, however, the twelve points of the original icosahedron have only five neighbours and are often called "pentagon points". It is stored following the recursive cuts ($k$) first and then the "levels" resulting from the original edge bisections. Thus, the storage approximates a space-filling curve locally. The choice of $n$ and $k$ can be arbitrary when generating a grid at a specific resolution.

The number of neighbouring grid cells around a central cell at cell-distance $r$ is: $N_{N(r)} = 12 \sum_{i=1}^{r} 6r(r+1)$. For example, $N_{N(1)} = 12$, $N_{N(2)} = 36$, $N_{N(3)} = 72$. This does not apply if pentagon points are within the neighbourhood.

The actual ICON grid is a result of an optimization process that aims to reduce the impact of the irregularity at the pentagon points by "stretching" the grid to an ideal sphere with two pentagon points placed exactly at the North, and South pole, respectively. This procedure slightly deforms the grid cells but does not change their number of neighbourhood relations and is thus not relevant for performance directly. Yet, the changed numeric properties may lead to a higher required resolution to resolve certain processes.

The grid is then instantiated into the third dimension by adding height coordinates on each cell extending it into a vertical column. Since clouds are moving in nonhydrostatic models, the pressure is not simply a function of the height but depends on the mass of the 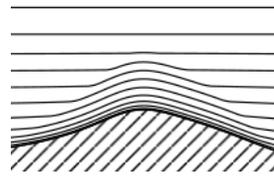
air column above, which is computed based on the atmospheric state. This is illustrated in the figure on the right. Thus, ICON uses height-based coordinates that follow the terrain. The total number of levels is, like the horizontal grid configuration, a system parameter, called num_lev.

### 3.3 ICON Time and Space Discretization

ICON simulates discrete time steps to advance the state of the grid. To solve the Euler equation, an approximation of the Navier Stokes equations, the dynamical core uses an explicit predictor-corrector solver for each time-step. In space, all operations are explicit except for the vertical discretization which is implicit and is solved using the Thomas algorithm. The vertical integration runs on each column independently.

With the explicit time stepping scheme and the Courant–Friedrichs–Lewy (CFL) condition, the time step must be small enough such that the fastest waves in the system (i.e., sound waves) do not cross over adjacent grid points to produce correct results. However, not all physical equations are bound to sound waves like the air fluid solver. For example, transport processes can be solved on slower timescales given wind speeds and air densities. Thus, ICON runs the dynamical core with smaller time steps than the tracer (usually 5x).

ICON generally fixes the transport process as the default time step $\delta t$ and defines all other time-steps relative. E.g., the dynamical core time step is then $\delta \tau = 5$ is defined as the frequency increase relative to $\delta t$. Some physical process parametrizations (e.g., radiation, convection, gravity wave drag) can live on even longer timescales and reduce the frequency with respect to $\delta t$. Those can be chosen individually for each physical process model, for example the expensive radiative transfer is often updated only every 30 minutes. There are some constraints, e.g., it is recommended that radiation and convection are called at the same steps.

As a rule of thumb, the maximum time step $\delta t$ allowed depends on the resolution: $\delta t < 9\ \delta x\ s/km$ (Prill et al, 2023) (assuming $\delta \tau = 5$). Furthermore, it is recommended to choose $\delta t$ always smaller than 1000s for numerical stability. $\delta t$ is set by the configuration variable dtime.

### 3.4 ICON Components

ICON contains multiple models, but the basic structure is always divided into three major components: dynamical core, numerical advection/tracer transport, and physical parameterization. The dynamical core solves the governing equations for fluid motion forward in time, the advection aka tracer transport scheme moves entities (e.g., humidity and clouds) according to the fluid motion, and the physics emulate processes that are happening on scales too small for the grid (e.g., cloud formation). We separate the transport from the dynamical core due to its slower time stepping.

The tracer transport works on a number of tracers, where each models a different physical entity. Tracers are essentially three-dimensional fields (arrays) that track certain entities (e.g., dust, hail, etc.). The number and type of tracers is configured by the user.

ICON offers different horizontal and vertical tracer implementations, and some may require smaller time-steps for numerical stability, called "sub-cycling". We model all tracers (specific to a physics process configuration) as a single entity.

The physics processes are called at each time step after the grid has been updated by the dynamical core, horizontal diffusion, and tracer transport. ICON differentiates between fast and slow physics. Each fast physics process is called at each (tracer) timestep, updates the variables, and passes the updated grid to the next fast physics process. Slow physics processes (e.g., cloud cover, radiation, gravity wave drags) are stepped forward in time independently as specified by the user. All physical processes are generally limited to one column and do not interact with neighbouring columns.

ICON output data can be generated on different grids and supports re-starting from a checkpoint. The user can specify an output time interval to determine the output frequency.

The components for any given configuration may be different and thus, the performance model will be specific to the configuration. Here, we provide a method to identify the different components such that they can be modelled independently to determine whether they become bottlenecks or not. While the dynamical core and the tracer are part of every configuration, the physical parameterizations are specific to each configuration. Below, we utilize two "aquaplanet" setups, one using the "AES" (Atmosphere Earth System) physics we received from the Max Planck Institute for Meteorology (MPI-M) and the German Climate Research Centre (DKRZ) as well as "NWP" physics (Numerical Weather Prediction) we received from the EXCLAIM team at ETH Zurich and MeteoSwiss.

We first relate the overview structure above to the detailed timers in the code, we name the timers in brackets (timer). We identify kernels of interest and assign short names in square brackets [name]. The dynamical core consists of three main components: the nonhydrostatic solver (nh_solve) and the nonhydrostatic diffusion operator (nh_diff) and the tracer (transport). We model the dynamical core with two phases [solv] for the substepped solver and [tran] for the diffusion and transport operators. The physics differ between the AES and NWP configurations.

The "Atmosphere Earth System" (AES) physical parameterization combines many different schemes. It also uses its own data representation and thus explicitly copies data from the dynamical core to the physics representation (dyn2phy, including boundary condition preparation (aes_bcs)) and back (phy2dyn) [pini]. In our current configuration, we use six physics parameterization schemes: (1) cloud cover (interface_aes_cov) [pcov], (2) radiation (interface_aes_rad) [prad], (3) radiative heating (interface_aes_rht) [prht], (4) vertical diffusion and surface (interface_aes_vdf) [pvdf], (5) (micro physics) graupel scheme (interface_cloud_mig) [pmig], and (6) WMO tropopause height (interface_aes_wmo) [ptro].

The call-sequence (and time tree) in the AES configuration is

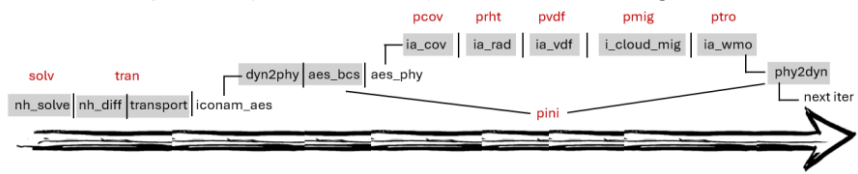

It shows one iteration of the time step loop, the associated icon timers, and our assigned phase names (in red - related to the grey-shaded timers). Vertical levels are call depth, e.g., iconam_aes invokes dyn2phy.

The "Numerical Weather Prediction" (NWP) physical parameterization uses several different schemes as well. The call-tree in the NWP configuration is

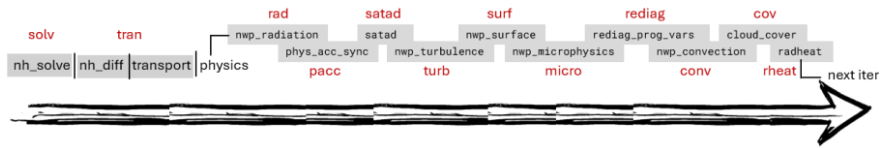

It shows one iteration of the time step loop, the associated icon timers, and our assigned phase names (in red - related to the grey-shaded timers). Here, we differentiate ten physics kernels.

We currently do not model the static grid refinement in ICON because this is largely used in regional setups. Furthermore, we do not explicitly consider physics parameterizations that are nor part of our setup. We also do not explicitly model physics executed on different grids, e.g., a reduced radiation grid to reduce radiation computation overhead up to 2x. Those reduced grids may be part of a configuration but are not parametrized in the model. All those are simple to model with our proposed methodology below.

### 3.5 ICON Structure and Optimizations

The horizontal icosahedral grid does not have an obvious ordering and is stored in a one-dimensional array with an index for each cell that locally approximates a space-filling curve. Most arrays represent centres of the triangular grid cells, but some represent edges or vertices. To store three-dimensional variables, the one-dimensional array is extended with a second dimension representing the vertical level. The 1d array representing the horizontal dimensions is split into chunks of configurable size nproma such that this

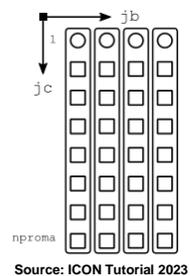

**Source: ICON Tutorial 2023**

1d array is stored in two dimensions and the overall array is stored in the order [horizontal index in block, vertical level, index of the block] as illustrated in the figure (Prill et al., 2023). This storage allows passing a full block including the vertical levels into a subroutine. Loops that iterate over grid cells, edges, or vertices are blocked in jb and jc for cache efficiency. The code uses indirect addressing to determine neighbourhood relationships.

ICON uses double precision per default and can be compiled to use mixed single and double precision calculations. Many local arrays in the dynamical core and some in tracer transport are then stored and computed as single precision variables.

### 3.6 ICON Parallelism

ICON offers OpenMP, OpenACC, and MPI parallelism. We first describe the MPI parallelization.

ICON can use limited functional parallelism to decouple logical steps: worker ranks advance the simulation, I/O ranks write output data, restart ranks write asynchronous restart data, and prefetch ranks read boundary data asynchronously. We will focus on a setup where we only use worker ranks and the simulation progress is halted during writing of the output data. The computational grid domain is distributed across the worker ranks.

The distribution is performed by cutting only the horizontal grid, i.e., each process owns all vertical levels of any triangular cell it owns. ICON uses a balanced recursive latitude-longitude bisection to determine which process owns which cell: given the whole (potentially partially refined or regional) grid, it cuts it first in two balanced longitude halves. Then, it proceeds to cut those halves each into two balanced latitude halves (for uneven process counts, assign the odd process to an arbitrary partition). The procedure is applied recursively until the number of processes is exhausted. Mapping grid cells to processes is flexible and one can think of other decompositions also including the vertical grid. Here, we describe what is used in the configurations we analysed.

The per-process grid is stored in the same arrays as in a sequential run. It distinguishes between inner cells and halo cells that are needed by neighbouring processes. All inner cells are stored first in the array and the halo cells are stored last.

### 3.7 Input problem - Aquaplanet

While a configuration defines how to combine the available modules into a simulation, the input problem determines the scale of the simulation (e.g., the resolution in space and time). An input problem defines the parameters of the performance model. We are interested in scalability with increased grid resolution as well as an increased number of MPI processes. The main critical performance parameters (Bauer et al., 2012) that are influenced by scaling these two entities are:

| Symbol | Description (name in configuration file) | Unit |
|---|---|---|
| $\delta t$ | the main (transport) time-step ("modelTimeStep") | s |
| $\delta rad$ | radiation time step ("aes_phy_config(1)%dt_rad") | s |
| $\delta vdf$ | vertical diffusion time step ("aes_phy_config(1)%dt_vdf") | s |
| $\delta mig$ | graupel scheme time step ("aes_phy_config(1)%dt_mig") | s |
| $\delta x$ | the effective resolution - specified by the grid refinement "RnBk", where $\delta x \approx 5050/(n \cdot 2^k)$ with the number of grid points $N = 20n^2 4^k$ | km |
| T | overall simulated time | s |
| P | number of worker-ranks in the MPI job | - |

When varying $\delta x$, we use the nine preconfigured grids[a] with the following $\delta t$ parameters:

| $\delta x$ [km] | n | k | N | max $\delta t$ [s] | actual $\delta t$ |
|---|---|---|---|---|---|

---

[a] from http://icon-downloads.mpimet.mpg.de/mpim_grids.xml

| | | | | | |
|---:|---:|---:|---:|---:|---:|
| 315.63 | 2 | 3  | 5,120       | 2,840 | 1800 |
| 157.81 | 2 | 4  | 20,480      | 1,420 | 900  |
| 78.91  | 2 | 5  | 81,920      | 710   | 450  |
| 39.45  | 2 | 6  | 327,680     | 355   | 225  |
| 19.73  | 2 | 7  | 1,310,720   | 177   | 150  |
| 9.86   | 2 | 8  | 5,242,880   | 88    | 75   |
| 4.93   | 2 | 9  | 20,971,520  | 44    | 25   |
| 2.47   | 2 | 10 | 83,886,080  | 22    | 20   |
| 1.23   | 2 | 11 | 335,544,320 | 11    | 10   |

We use $\delta rad$=1,800s (30 mins), $\delta vdf$=$\delta$t, $\delta mig$=$\delta$t (same as fast stepping), and nlev=70 for all runs and adjust the used $\delta t$ to be smaller than the maximum suggested and divide the physics time steps.

After some simple manipulation of the equations for $N$ and $\delta x$, we find $N = 20 \cdot 5050^2 / \delta x^2$.

We model the requirements of each invocation of each phase in isolation as $R_{phase}(\delta x, P = 1) = a_{phase}/(\delta x^2 P) + b_{phase}$ where $a_{phase}$ and $b_{phase}$ are phase-specific constants to express its requirements per process. For sequential models, the optional parameter $P$ is set to 1 and can be omitted. For example, for AES, $phase = \{solv, tran, pini, pcov, prht, pvdf, pmig, ptro\}$.

Now, for AES, we can then compute the overall requirements of an ICON run for a time-interval $T$ as:

$R(\delta x, T, P = 1) = T/\delta t \cdot (5\, R_{solv}(\delta x, P) + R_{tran}(\delta x, P) +$
$R_{pini}(\delta x, P) + R_{pcov}(\delta x, P) + \delta t/\delta rad \cdot R_{prad}(\delta x, P) +$
$R_{prht}(\delta x, P) + \delta t/\delta vdf \cdot R_{pvdf}(\delta x, P) +$
$\delta t/\delta mig \cdot R_{pmig}(\delta x, P) + R_{ptro}(\delta x, P))$
with $\delta t < 9\delta x$.

For the NWP configuration, we use the same $\delta t$ and $\delta x$ steps, but we vary the physical parameterization frequencies in step with $\delta t$: $\delta rad = 4 \cdot \delta t$, $\delta cov = 2 \cdot \delta t$, and $\delta conv = 2 \cdot \delta t$. For both the AES and NWP configurations we halve T every time we decrease $\delta t$ thus ensuring we are always simulating the same number of time steps:

$R(\delta x, T, P = 1) = T/\delta t \cdot (5\,R_{solv}(\delta x, P) + R_{tran}(\delta x, P) + \delta t/\delta conv\, R_{conv}(\delta x, P) + \delta t/\delta cov\, R_{cov}(\delta x, P) + \delta t/\delta rad \cdot R_{rad}(\delta x, P) + 2R_{turb}(\delta x, P) + R_{micro}(\delta x, P) + R_{satad}(\delta x, P) + R_{surf}(\delta x, P) + R_{rediag}(\delta x, P) + R_{radheat}(\delta x, P) + R_{accum}(\delta x, P))$
with $\delta t < 9\delta x$.

Thus, instantiating the rates in the general equation above, the requirements equation becomes:

$R(\delta x, T, P = 1) = T/\delta t \cdot (5\,R_{solv}(\delta x, P) + R_{tran}(\delta x, P) + 0.5\,R_{conv}(\delta x, P) + 0.5\,R_{cov}(\delta x, P) + 0.25 \cdot R_{rad}(\delta x, P) + 2R_{turb}(\delta x, P) + R_{micro}(\delta x, P) + R_{satad}(\delta x, P) + R_{surf}(\delta x, P) + R_{rediag}(\delta x, P) + R_{radheat}(\delta x, P) + R_{accum}(\delta x, P))$
with $\delta t < 9\delta x$.

We note that when using NWP, some physics components are also called once during initialization, therefore an additional initialization requirement exists:

$R_{init}(\delta x, P = 1) = R_{conv}(\delta x, P) + R_{cov}(\delta x, P) + R_{rad}(\delta x, P) + R_{satad}(\delta x, P) + R_{turb}(\delta x, P) + R_{radheat}(\delta x, P) + R_{accum}(\delta x, P)$

For long enough simulation times, $R_{init}$ should become negligible.

**3.8 Requirements modelling**

Our first models will determine the requirements of the phases of ICON, most of those are independent of the specific architecture. For example, the number of required floating point operations, the required memory, or communication volume solely depend on the configuration, input problem, and the domain decomposition. Others, such as cache misses, depend on the architecture (cache size).

In the following, we will utilize hardware counters to **count the machine-independent requirements** for several executions with varying critical parameters. For this, we compile ICON version rc2.6.7 with the GNU compiler suite version 10.2.0 with the optimization flag "-O2".

### *3.8.1  Required operation counts (sequential work)*

To collect performance counters, we extend ICON's timer infrastructure with LibLSB (Hoefler and Belli, 2023) with a small patch (<50 lines of code). We then count the total operations and parametrize the model: $R^{op}_{phase}(\delta x, P) = a^{op}_{phase}/(\delta x^2 P)$ and we also count the floating point operations and parametrize the model: $R^{fp}_{phase}(\delta x, P) = a^{fp}_{phase}/(\delta x^2 P)$.

When compiled in mixed precision, only the dyc phase uses fp32. In that case, 14-15% of the dyc flops are fp32, all others are fp64. The actual coefficients for $a^{fp}$ and $a^{op}$ for AES are shown in the following table:

| phase | $a^{fp}$ [Tflop] | $a^{op}$ [Top] | flop/op ratio |
|---|---:|---:|---:|
| solv | 19.23 | 811.37 | 2.37% |
| tran | 58.00 | 1,444.53 | 4.02% |
| pini | 8.43 | 255.73 | 3.30% |
| pcov | 0.12 | 5.78 | 2.03% |
| prad | 1,491.12 | 18,301.39 | 8.15% |
| prht | 0.37 | 14.63 | 2.55% |
| pvdf | 34.13 | 828.85 | 4.12% |
| pmig | 5.94 | 96.11 | 6.18% |
| ptro | 3.50 | 14.98 | 23.37% |

For NWP, the coefficients are

| phase | $a^{fp}$ [Tflop] | $a^{op}$ [Top] | flop/op ratio |
|---|---:|---:|---:|
| solv | 16.64 | 804.36 | 2.07% |
| tran | 43.14 | 986.63 | 4.37% |
| cov | 5.09 | 44.89 | 11.34% |
| rad | 705.40 | 9,691.36 | 7.28% |
| conv | 11.17 | 171.78 | 6.50% |
| micro | 7.42 | 60.81 | 12.20% |
| satad | 2.77 | 57.71 | 4.80% |
| turb | 10.51 | 185.31 | 5.67% |
| phys_acc_sync | 2.42 | 66.32 | 3.65% |
| surface | 0.01 | 0.23 | 6.46% |
| rediag | 3.13 | 37.12 | 8.43% |
| radheat | 3.41 | 24.36 | 14.00% |

The quality of all fits is with an $R^2$ of > 0.998 and thus very accurate. The table also shows the ratio of total operations to floating point operations for the different phases. Note that the floating-point operations are likely to be bound to the algorithm while the total operations depend on the target architecture as well as the compiler.

The cost of the radiation schemes in AES and NWP differs by more than a factor of two, which is largely due to the differences in the schemes and different implementations. The slight difference in the cost of solver (dynamical core) is likely due to the difference in the way the Rayleigh damping near the model top is used. The AES configuration has a larger number of vertical levels in the damping layer thereby adding to the extra cost.

This model is already useful to understand compute requirements for various configurations. For example, let us model the requirements for a run for 30 years (946M seconds) with 1.2 km resolution and our aquaplanet configuration from above. We choose $\delta t = 10s$, which makes 94.6M transport time steps and 437M dycore time steps. Se we can parametrize the total flop required equation to:

$$R(1.2, 946M) = 94.6M \cdot (5\, R_{dyn}(\delta x) + R_{tran}(\delta x) + R_{pini}(\delta x) + R_{pcov}(\delta x) + 1/180 \cdot R_{prad}(\delta x) + R_{prht}(\delta x) + 1/90 \cdot R_{pvdf}(\delta x) + 1/90 \cdot R_{pmig}(\delta x) + R_{ptro}(\delta x))$$

Resulting in the requirements for AES are listed in the following table (in Zettaops, i.e., 1e21 ops)

| phase | Zflop | Zop | share of total |
|---|---:|---:|---:|
| solv | 5.9857 | 252.4956 | 54.85% |
| tran | 3.6100 | 89.9067 | 33.08% |
| pini | 0.5246 | 15.9166 | 4.81% |
| pcov | 0.0073 | 0.3597 | 0.07% |
| prad | 0.5156 | 6.3281 | 4.73% |
| prht | 0.0232 | 0.9105 | 0.21% |
| pvdf | 0.0236 | 0.5732 | 0.22% |
| pmig | 0.0041 | 0.0665 | 0.04% |
| ptro | 0.2178 | 0.9322 | 2.00% |
| **sum** | **10.9119** | **367.4891** | **100.00%** |

For NWP, we get

| phase | Zflop | Zop | share of total |
|---|---:|---:|---:|
| solv | 5.1774 | 250.3140 | 48.76% |
| tran | 2.6852 | 61.4073 | 25.29% |
| cov | 0.0035 | 0.0310 | 0.03% |
| rad | 0.2439 | 3.3510 | 2.30% |
| conv | 0.0077 | 0.1188 | 0.07% |
| micro | 0.4619 | 3.7845 | 4.35% |
| satad | 0.1724 | 3.5920 | 1.62% |
| turb | 1.3084 | 23.0668 | 12.32% |
| phys_acc_sync | 0.1508 | 4.1275 | 1.42% |
| surface | 0.0009 | 0.0140 | 0.01% |
| rediag | 0.1947 | 2.3106 | 1.83% |
| radheat | 0.2123 | 1.5162 | 2.00% |
| **sum** | **10.6191** | **353.6338** | **100.00%** |

### *3.8.2 Communication Characteristics*

For communication, we consider three increasingly complex aspects: (1) the communication pattern, (2) the total communication volume per phase, and (3) the message count and size distribution. We collect the data using liballprof (a part of the LogGOPSim toolchain (Hoefler et al., 2010)) to profile each call to any MPI function during parallel program runs and post-process those using bespoke Python scripts.

We first illustrate the decomposition and messaging on an idealized square domain with periodic boundary conditions in both dimensions (the triangular grid will be similar).

For a $\sqrt{N}x\sqrt{N}$ square domain, we see that for the special case with two processes, there is only one message of size $2\sqrt{N}$ (left and right boundary) exchanged. Beginning with four processes, each process exchanges the top and bottom as well as the left and right boundary with four other processes; the size of each exchange is $\sqrt{N}/2$. For eight processes, each process exchanges two messages of size $\sqrt{N}/2$ for top and bottom and two messages of size $\sqrt{N}/4$ for left and right. For sixteen processes, each process exchanges messages of size $\sqrt{N/4}$ with four neighbors.

We only consider powers-of-two process-counts beginning from four processes in our modelling. If P is a power of four, each process sends four messages of size $\sqrt{N}/\sqrt{P}$ to each of the neighbors. If P is a power of two but not a power of four, then each process sends two messages of size $\sqrt{N}/\sqrt{2P}$ and two messages of size $2\sqrt{N}/\sqrt{2P}$ to four neighbors.

Thus, each process has exactly four faces for $P \geq 16$ due to the decomposition symmetry. The actual number of communication partners varies between 4 and 10 across different ranks. This is due to the details of the distributed halo zones and the icosahedral grid where most elements have six neighbours but some elements have five neighbours.

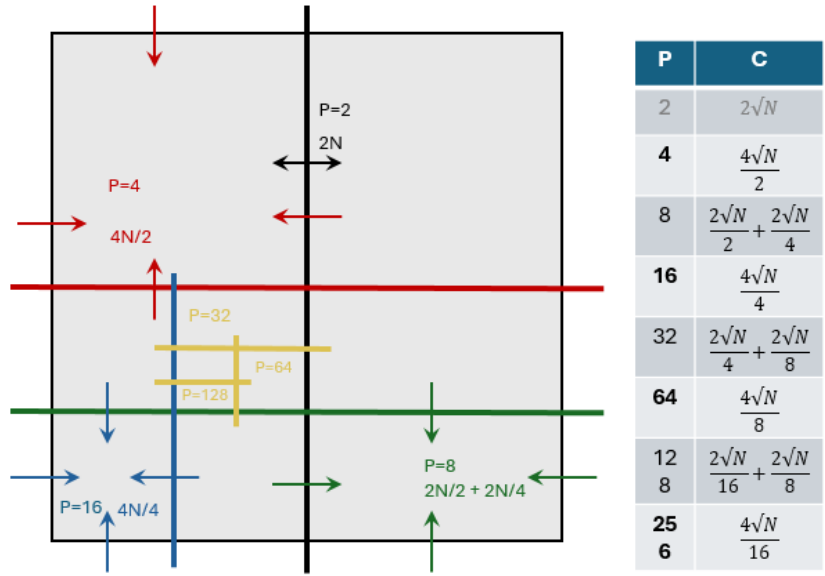

| P | C |
|---|---|
| 2 | $2\sqrt{N}$ |
| 4 | $\frac{4\sqrt{N}}{2}$ |
| 8 | $\frac{2\sqrt{N}}{2} + \frac{2\sqrt{N}}{4}$ |
| 16 | $\frac{4\sqrt{N}}{4}$ |
| 32 | $\frac{2\sqrt{N}}{4} + \frac{2\sqrt{N}}{8}$ |
| 64 | $\frac{4\sqrt{N}}{8}$ |
| 128 | $\frac{2\sqrt{N}}{16} + \frac{2\sqrt{N}}{8}$ |
| 256 | $\frac{4\sqrt{N}}{16}$ |

Since N depends on $\delta x$, we model the communication volume per process per phase as:

$C_{phase}(\delta x, P) = a^{com}{}_{phase}/(\delta x \cdot \sqrt{P}) + b^{com}{}_{phase} \cdot \sqrt{P}$ for $P$ being a power of four where $a^{com}{}_{phase}$, $b^{com}{}_{phase}$ are phase-specific constants to express its requirements per process.

In NWP, only the solv, tran, and one physics parameterization phase (phys_acc_sync)) communicate. We model the communication volume for those and provide the coefficients in the table below:

| phase | $a^{com}$ [MB] | $b^{com}$ [KB] |
|---|---:|---:|
| solv | 757.20 | 0.2918 |
| tran | 1375.77 | 0.4176 |
| phys | 949.77 | 0.2803 |

Here again, the $R^2$ is bigger than 0.998 indicating an excellent fit. For the 30-year simulation with a resolution of 1.23 km, the following communication volumes would be expected:

| phase | $a^{com}$ [PB] | $b^{com}$ [TB] | share of total |
|---|---:|---:|---:|
| solv | 2.27 | 3.53 | 61.84% |
| tran | 0.82 | 5.06 | 22.58% |
| phys | 0.57 | 3.39 | 15.58% |
| **sum** | **3.66** | **11.98** | **100.00%** |

The message size distribution is more complex, and we will model only bounds and show distributions empirically. We show examples for a fixed grid and varying numbers of processes (strong scaling) as well as a varying grid and fixed number of processes.

We first vary the grid size, while keeping the number of processes constant (P=8). The results show in order R2B3 (top left), R2B4 (top right), R2B5 (bottom left), R2B6 (bottom right). We notice that the distribution remains similar, but the size of messages being exchanged roughly doubles with each increase in grid size.

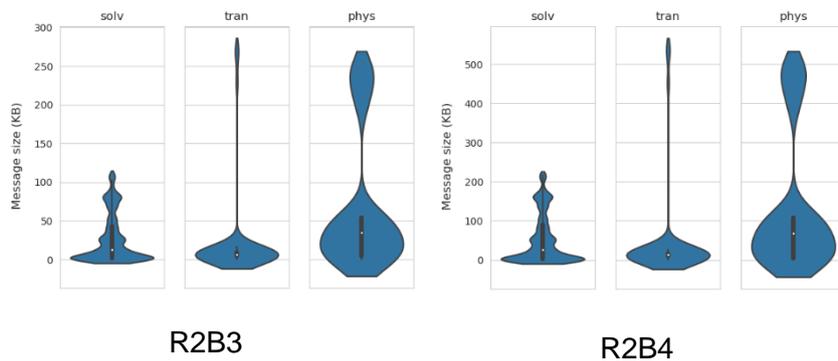

R2B3　　　　　　　　　　　　R2B4

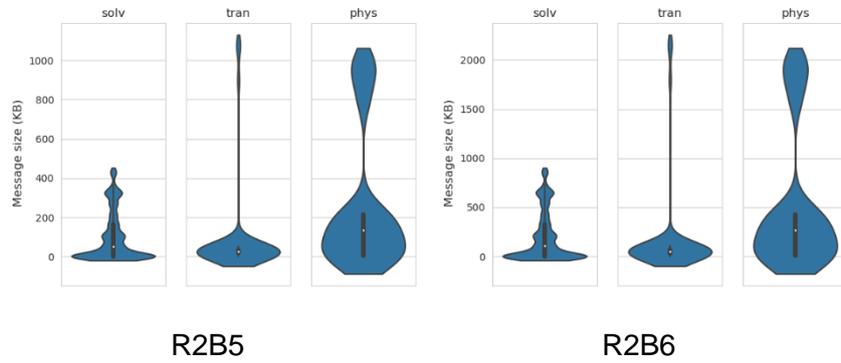

R2B5           R2B6

The same cannot be said about the variation of MPI processes, which has a more complex impact on the distribution of message sizes and their counts and cannot be easily fit to an analytical expression. This is also a result of the icosahedral (triangle) decomposition in which the neighbourhood relations differ across processes. In the following, we keep the grid size constant (R2B3) and vary the number of MPI processes. The results show in order 2 processes (top left), 4 processes (top right), 8 processes (middle right), 16 processes (middle left), 32 processes (bottom left), 1024 processes (bottom right):

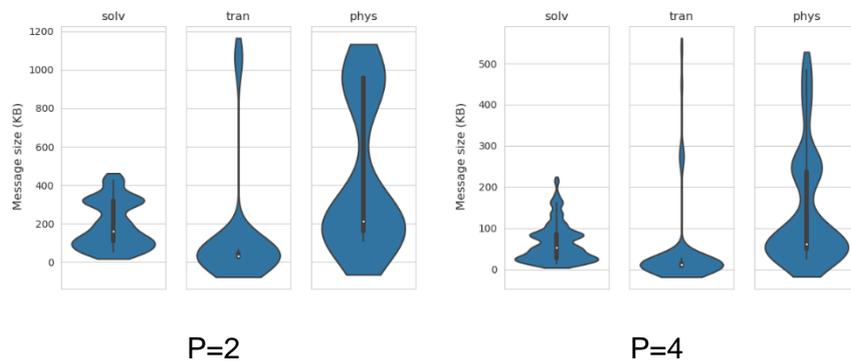

P=2           P=4

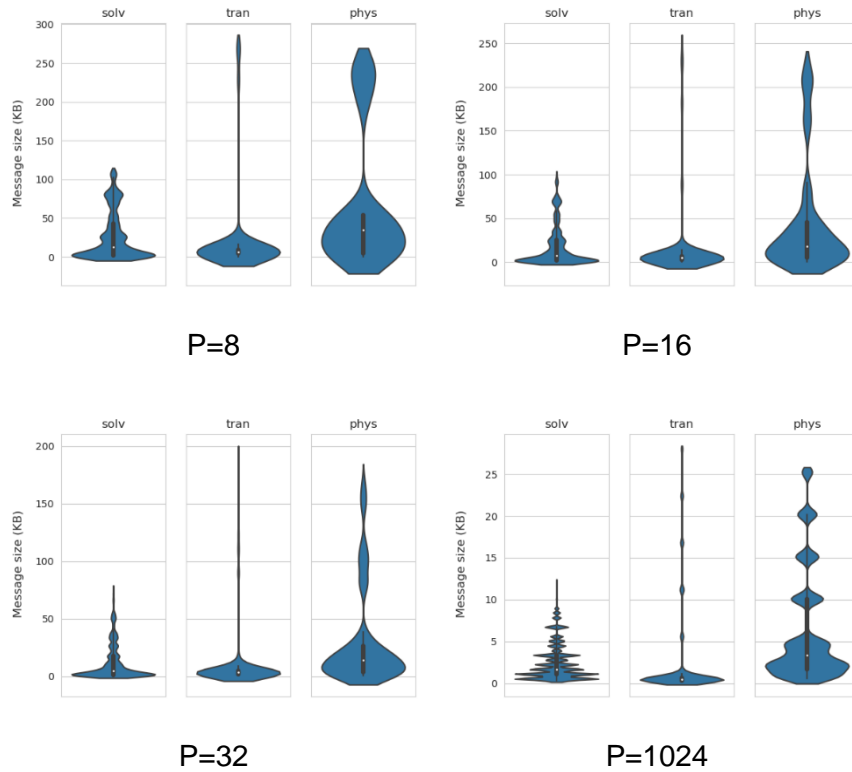

### 3.8.3 File I/O volume

ICON can write the simulation data in multiple formats to disk. For the output, one defines a variable list and output intervals for each variable. The output size is simply the number of written elements per step E times the number of output time steps per variable $T_o$. The number of points per level is $N = 20 \cdot 5050^2 / \delta x^2$ and $E = N \cdot nlev$.

For example, for a 30-year simulation at approximately 1km resolution with an output timestep of 15 minutes and values in FP16, the amount of storage needed per 2D variable is 705 TiB and per 3D variable with 70 levels would be 49 PiB. Usually, different variables are written at different intervals. For example, wind-speed requires high-frequency output while temperature or humidity can often be recorded at lower frequencies.

### *3.8.4 Memory requirements*

We also consider maximum memory requirements as a machine to execute ICON must have at least this amount of memory available for a run. For this, we measured the maximum memory consumption (VmPeak in Linux) through various run configurations depending on N ($\delta x$) and P.

As expected, the memory consumption per MPI rank grows with the grid and decreases with an increasing number of MPI ranks (modelled by $a$). We also observe a term in the memory consumption that grows with the number of MPI ranks (modeled by $b$), explained by the local data structures required by each MPI process that depend on the total number of processes as well as sizable constant memory requirement (modeled by $c$). The memory requirement M can therefore be expressed as follows: $M(\delta x, P) = a_M/(\delta x^2 P) + b_M \cdot P + c_M$. In our configuration, we found the following concrete values: $M(\delta x, P) = 20 \cdot 5050^2 \cdot 1.01/(\delta x^2 P) + 4.6 \cdot P + 915.6$ [MiB].

### 3.9 Summary and Conclusions

Our performance model enables scientists to fully assess requirements for climate simulations with the ICON code. It not only counts the number of required floating point operations accurately, but it also counts messaging volume and I/O volumes. The relative magnitude of these requirements will allow application users to configure each simulation run to a particular system and it will also allow system designers to optimize future supercomputers to the ICON weather and climate workloads.

Specifically, our model allows scientists to reason about the computational requirements for the next step in global km-scale climate simulations.

## Acknowledgements

We thank W. Sawyer, T. Jahns, F. Prill, and B. Stevens for useful discussions that improved the manuscript.